\documentclass[%
 reprint,
 amsmath,amssymb,
 aps,
]{revtex4-1}

\usepackage{graphicx}
\usepackage{dcolumn}
\usepackage{bm}

\begin{document}

\preprint{APS/123-QED}

\title{Insight of the thermal conductivity of $\epsilon-$iron at Earth's core conditions from the newly developed direct $ab~initio$ methodology}

\author{Sheng-Ying Yue}
\email{sheng-ying_yue@ucsb.edu}
\affiliation{Aachen Institute for Advanced Study in Computational Engineering Science (AICES),
RWTH Aachen University, 52062 Aachen, Germany}
\affiliation{Department of Mechanical Engineering, University of California, Santa Barbara, CA 93106, USA}

\author{Ming Hu}
\email{HU@sc.edu}
\affiliation{Aachen Institute for Advanced Study in Computational Engineering Science (AICES),
RWTH Aachen University, 52062 Aachen, Germany}
\affiliation{Institute of Mineral Engineering, Division of Materials Science and Engineering, Faculty of Georesources and Materials Engineering, RWTH Aachen University, 52064 Aachen, Germany.}
\affiliation{Department of Mechanical Engineering, University of South Carolina, Columbia, SC 29208, United States.}

\date{\today}

\begin{abstract}
The electronic thermal conductivity of iron at Earth's core conditions is an extremely important physical property in geophysics field. However, the exact value of electronic thermal conductivity of iron under extreme pressure and temperature still remains poorly known both experimentally and theoretically. 
A few recent experimental studies measured the value of the electronic thermal conductivity directly and some theoretical works have predicted the electronic thermal conductivity of iron at Earth's core conditions based on the Kubo-Greenwood method. However, these results differ largely with each other. A very recent research has confirmed that for iron at Earth's core conditions the strength of electron-electron scattering could be comparable to that for electron-phonon scattering, 
meaning that the electron-electron scattering should also be considered when evaluating the electronic thermal conductivity in the Earth's core situations. Here, by utilizing a newly developed methodology based on direct non-equilibrium $ab~initio$ molecular dynamics simulation coupled with the concept of electrostatic potential oscillation, we predict the electronic thermal conductivity of iron in h.c.p phase. Our methodology inherently includes the electron-phonon and electron-electron interactions under extreme conditions. Our results are comparable to the previous theoretical and experimental studies. More importantly, our methodology provides a new physical picture to describe the heat transfer process in $\epsilon-$iron at Earth's core conditions from the electrostatic potential oscillation point of view and offers a new approach to study thermal transport property of pure metals in planet's cores with different temperature and pressures.
\end{abstract}

\maketitle


\section{Introduction}
\label{sec:intro}
The decay of radiogenic isotopes and slow cooling process are one of reasons to drive the gigantic thermal engine of the Earth. The geological dynamics induces mountain building, volcanoes and plate tectonics, etc. Another very important result of the geological dynamics is the geomagnetic field, which is generated in the liquid iron core via a thermal dynamo mechanism in the Earth's core cooling and freezing process\citep{Pozzo-nature}. 
A recent review article by Williams\citep{williams-2018} concluded that, the thermal conductivity of iron alloys under the extreme temperature and pressure is a very critical parameter for the heat flow out of Earth's core at present. It is also important for the inferred age of Earth's inner core. It will also help us understand the thermal evolution of Earth's core and lowermost mantle.
Thus, it is significant to investigate the thermal transport property and the dynamical evolution of the Earth's core. As we all know, the Earth's inner-solid core is mainly composed of metal iron (Fe)\citep{Tateno359}. For example, in the previous experimental measurements, the isothermal equation of state of $\epsilon-$iron (hcp phase) and $\gamma-$iron (fcc phase) indicates that the inner core is mainly composed of pure solid iron\citep{Anderson}. It was further confirmed that the Earth's core consists of a molten iron alloy outer core surrounding a solid inner-core of $\epsilon-$iron\citep{Seagle}. Previous $ab~initio$ calculations\citep{Pozzo-EPSL,Pourovskii} predict a high thermal conductivity for $\epsilon-$iron generally assumed to form the inner core. Although some other researches showed that the Earth's inner core is not pure Fe\citep{Fe-phase-1,Fe-phase-2,Fe-phase-3,Fe-phase-4}, from fundamental research point of view, it is still significant to study the pure-phase Fe at the Earth's core conditions.
Despite the efficiency of heat transfer process in Earth's core determines the dynamics of convection and limits the power available for the geodynamo directly\citep{Seagle}, prediction of heat transfer of $\epsilon-$iron under the extreme Earth's core conditions, most critically the electronic and lattice thermal conductivities, still remain unclear because of intrinsic experimental and theoretical difficulties\citep{Ross}.
For instance, in 2016, the experimental studies from two research groups of Ohta et al.\citep{Ohta} and Konopkova et al.\citep{Konopkova} present opposite results for the thermal conductivity.

\section{Theory}
Taking theoretical investigation as an example, the previous theoretical studies adopted the Kubo-Greenwood method to evaluate $\kappa_{el}$ of iron at the Earth's core conditions\citep{Pozzo-nature, Koker-pnas, Pozzo-EPSL, Pourovskii}. However, these studies adopted the Kubo-Greenwood formula as starting point for evaluation of kinetic coefficients, and in this process the electron-electron interaction is not included\citep{Seagle}.
In addition, a theoretical work shows that the $\epsilon-$iron is found to behave as a nearly perfect Fermi liquid and the quadratic dependence of the scattering rate in Fermi liquids leads to a modification of the Wiedemann-Franz law (WFL) with suppression of the $\kappa_{el}$ as compared to the electrical one. The strongly correlated electron system significantly increases the electron-electron thermal resistivity, which is found to be of comparable magnitude to that induced by the electron-phonon interaction\citep{Pourovskii}. This is why we need to consider the electron-electron scattering when evaluating the $\kappa_{el}$ at the Earth's core conditions. 

\begin{figure}[ht!]
\centerline{\includegraphics[width=1.0\linewidth,clip]{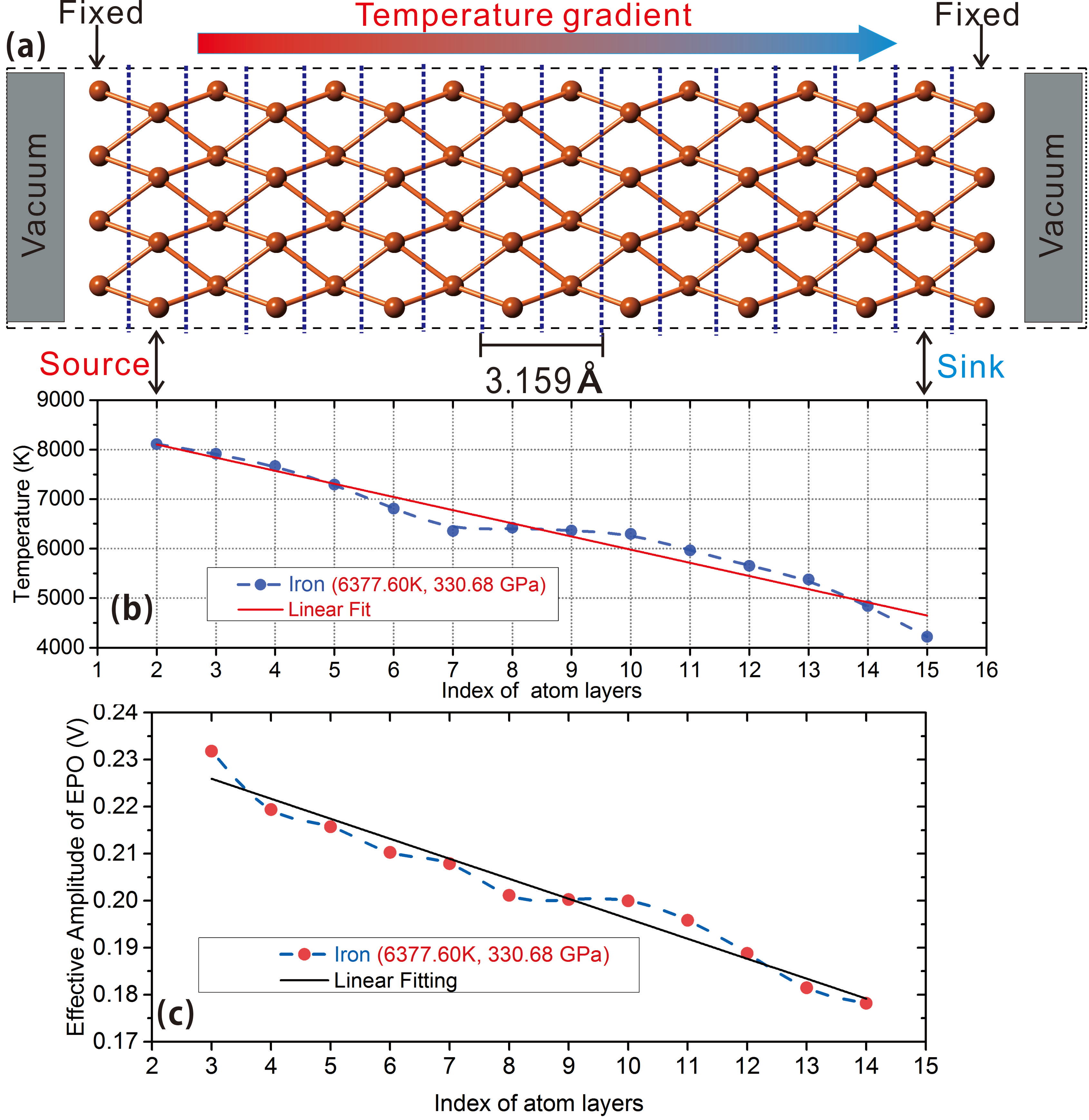}}
\caption{\textbf{(a)} illustrates the simulation model ($\epsilon-$iron) at inner core boundary (ICB) conditions used in our NEAIMD simulations. Fixed boundary conditions at both ends are adopted and the vacuum layers outside of the fixed layers are set up. The periodic boundary conditions are applied in the two lateral directions. \textbf{(b)} is the temperature profile from a representative NEAIMD simulation. \textbf{(c)} is the effective amplitude of electrostatic potential oscillation (EPO) from NEAIMD along the heat flux direction. Here, the representative NEAIMD simulation of $\epsilon-$iron is at 6377.60\,K and 330.68\,GPa and simulation time is 9~ps.}
\label{fig:1}
\end{figure}

Recently, we developed a new methodology based on direct non-equilibrium $ab~initio$ molecular dynamics (NEAIMD) calculations, which can simultaneously predict the electronic thermal conductivity and lattice (phonon) thermal conductivity of pure metals under the first-principles framework\citep{Yue}. Here, we apply our newly developed methodology\citep{Yue} to evaluate the electronic ($\kappa_{el}$) and phononic ($\kappa_{ph}$) thermal conductivities of $\epsilon-$iron at the inner core boundary (ICB) and the core mantle boundary (CMB). Our methodology can directly predict the electronic and phononic thermal conductivities of pure iron metal at the Earth's core conditions with high accuracy, without necessity of explicitly addressing any complicated scattering process of free electrons\citep{Yue}. This method also provides a new physical picture to describe the electronic thermal transport. Because the NEAIMD we adopt here is based on the Born-Oppenheimer approximation, it means that during every molecular dynamics step the electronic wave functions will have self-consistent calculation based on the Kohn-Sham equations to reach the convergence criteria. In this process the electron-electron interactions are naturally included, which is one of the major advantages of our NEAIMD method\citep{Yue}. Here, our simulation results as shown below are comparable to some previous theoretical studies\citep{Pozzo-EPSL,Davies-nature, Pourovskii, Secco}. It is shown that the thermal conductivity of $\epsilon-$iron in the Earth's core is high, which corresponds to a recent experimental result\citep{Ohta}. This can support the point of view that the Earth's core rapid cooling leads to an inner core younger than 0.7 billion years\citep{Ohta,Labrosse}. Then, a geophysical inference can be obtained that the abrupt increase in palaeomagnetic field intensity 1.3 billion years ago\citep{Ohta,Biggin} may not be related to the Earth's core birth.

Moreover, different from the conventional WFL theory, the methodology we referred herein describes the electronic thermal transport with a new physical concept. The vibrations of ions will induce the electrostatic potential oscillation (EPO) in space. The physical picture we propose is that, the local EPO can drive the valance shell electrons to collectively oscillate. Then, in the momentum space, the free electrons which are near the Fermi surface could be excited above the Fermi surface and gain the additional thermal kinetic energy simultaneously. Higher temperature corresponds to larger and faster atomic vibration, which will lead to stronger EPO in space. Then, the thermally excited electrons (TEEs) from the high-temperature region will possess more kinetic energy than those in the low-temperature region. When a stable distribution of the thermal kinetic energy of TEEs in space is established along $\nabla \vec{T}$ (temperature gradient), the heat flux carried by TEEs $\vec{J}_{el}$ and $\kappa_{el}$ can be calculated eventually\citep{Yue}.

In principle, this methodology includes all the interactions and scattering process of electrons, such as complex electron-electron interactions and electron-phonon interactions. This method does not depend on any artificial parameters and approximations. Thus, it can provide accurate results of $\kappa_{el}$ of pure metals. In addition, the non-equilibrium $ab~initio$ molecular dynamics-electrostatic potential oscillation (NEAIMD-EPO) method can predict the phononic thermal conductivity ($\kappa_{ph}$) of metals simultaneously. Therefore, the NEAIMD-EPO method provides us full access to evaluate the heat transfer process of $\epsilon-$iron in Earth's core conditions. More specific details of the methodology can be found in our recent paper\citep{Yue}.

\begin{figure}[th]
\centerline{\includegraphics[width=1.0\linewidth,clip]{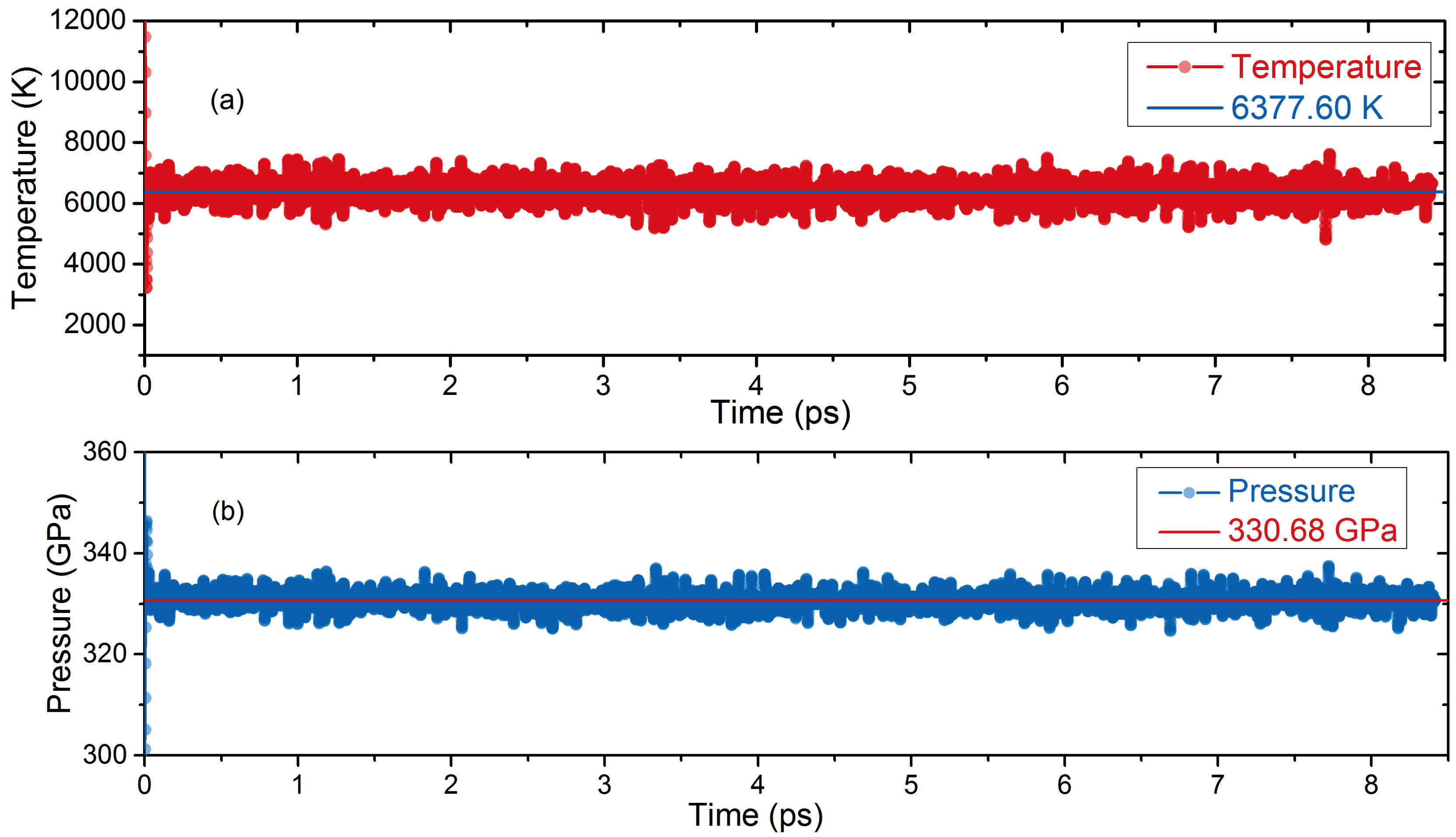}}
\caption{System temperature \textbf{(a)} and pressure \textbf{(b)} oscillate with respect to time for a representative NEAIMD simulation of $\epsilon-$iron at the inner-core boundary (ICB) conditions.}
\label{fig:2}
\end{figure}

\section{Computational Methods}
We perform NEAIMD simulations through the modified the Vienna $ab~initio$ Simulation Package (VASP)\citep{Stephen, vasp01, vasp02}, with the Perdew-Wang (PW91) functional\citep{PW91-01, PW91-02} and the projector augmented wave (PAW) method\citep{PAW01,PAW02}. The electronic exchange-correlation is treated using the generalized-gradient approximation (GGA) of Perdew et al.\citep{GGA}. In achieving good accuracy for iron at high-pressures, the choice of orbitals that can be treated as valence orbitals is crucial\citep{Gilles}. At ambient pressure, reasonable accuracy can be obtained by treating all atomic states up to and including $3p$ as core states, but this is not satisfactory here because the $3p$ states respond significantly at high pressure\citep{Gilles}. This is mainly because of the fact that $\epsilon-$iron under extreme conditions behaves as a Fermi liquid\citep{Pourovskii}. In contrast, We have tested the effect on the conductivity of the inclusion in valence of semi-core $3s$ and $3p$ states. We found that inclusion of $3s$ states in the atomic core gives insignificant errors, provided that the nonlinear core corrections are included\citep{Gilles,3s-error}. Thus, the $3p^6 4s^1 3d^7$ valence electronic configuration is adopted in this study and the core radii is $\rm 1.16~\overset{\circ}{A}$\citep{Gilles, Pozzo-EPSL, Pozzo-nature}. The single atomic orbitals are expanded in plane-waves with an energy cutoff of $\rm 400~eV$. The electronic energy levels are occupied according to the Fermi-Dirac statistics, with electronic temperature corresponding to the real system temperature (two temperature model). All the calculation details are provided in Sec.~1 of Supplementary Information (SI). To examine the correctness of our strategy for adopting the pseudo-potential, we also calculate the thermal conductivity of b.c.c iron at ambient condition. The total thermal conductivity (electronic + phononic) of b.c.c iron is obtained to be $\rm 83.68~W/(mK)$ at $\rm 296~K$, which is in very well agreement with the experimental data of $\rm 80.2~W/(mK)$ at $\rm 300~K$ and the previous theoretical research\citep{RE-Cohen,bcc-Alfe}. The b.c.c iron results are presented in Sec.~2.6 of SI.

An efficient extrapolation of the charge density was used to speed up the NEAIMD simulations\citep{Alfe} by sampling the Brillouin zone (BZ) with the $\Gamma$ point only. The temperature was controlled with Andersen thermostat\citep{Andersen-Thermostat} and the timestep was set as $\rm 1~fs$. We run simulations for typically $\rm 8-10~ps$, from which we discarded the first $\rm 0.3~ps$ to allow the system to reach equilibrium. We realize the atomic heat flux in NEAIMD via the M\"uller-Plathe algorithm\citep{Muller-Plathe}. In the M\"uller-Plathe method, the kinetic energies of the atoms (nuclei) in the heat source and heat sink are continuously exchanged. After sufficient simulation time, we can establish a relatively steady temperature gradient in metals. Figs.~\ref{fig:1}(a,~b) exhibit a representative $\epsilon-$iron model ($4\times4\times8$ supercell including 256 atoms) for NEAIMD-EPO simulation and the corresponding temperature profile under the ICB (6377.60~K, 330.68~GPa) condition, respectively. An animation of iron atom movement for a representative NEAIMD-EPO simulation is also provided as Supplemental Materials.

\begin{figure}[th]
\includegraphics[width=1.0\linewidth,clip]{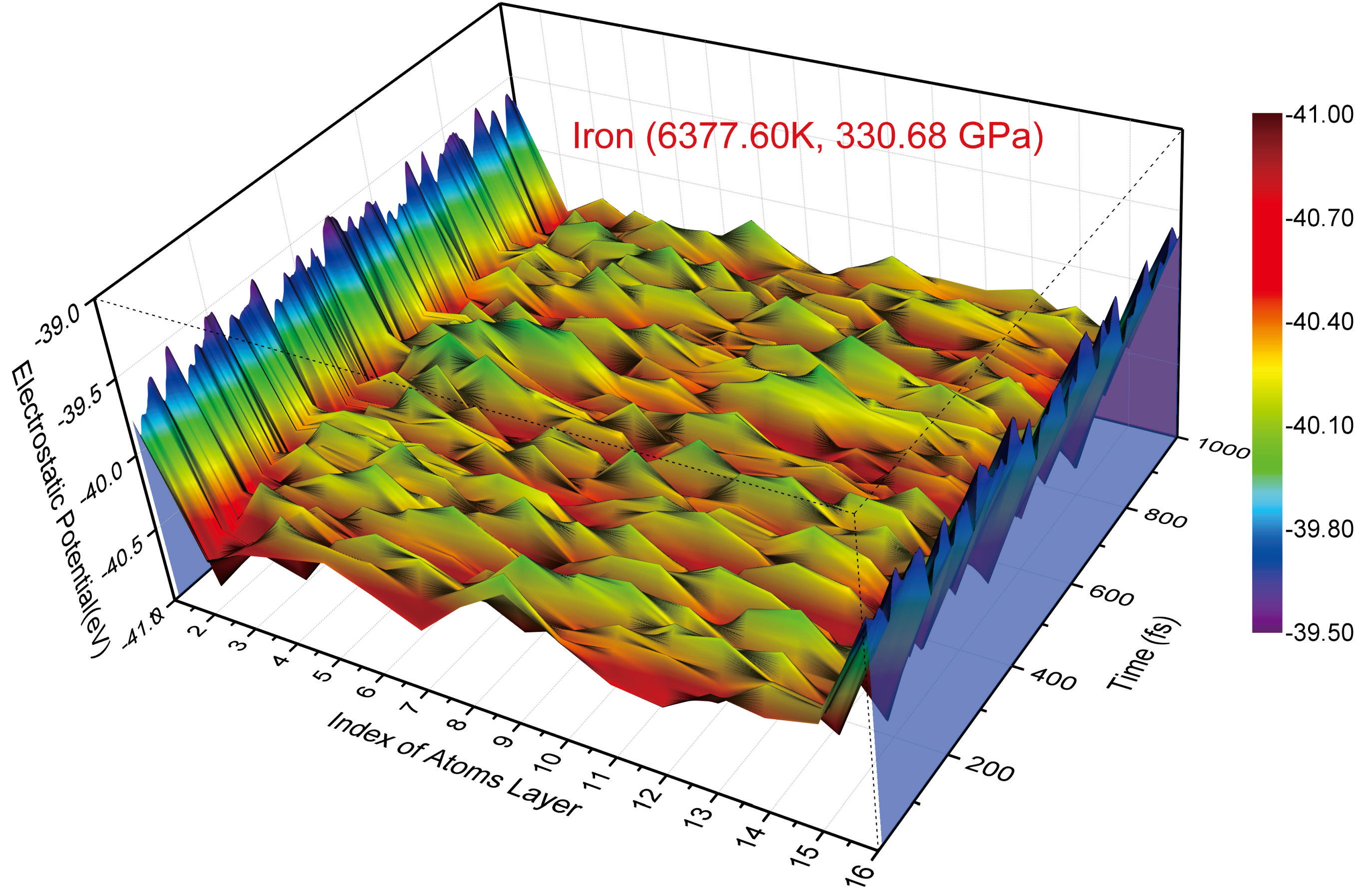}
\caption{The variation of electrostatic potential oscillation (EPO) in space with simulation time (where we normalize the test charge number to 1). The plotting data originates from the 9-ps NEAIMD-EPO simulation of $\epsilon-$iron at the inner-core boundary (ICB) condition (6377.60~K, 330.68~GPa).}
\label{fig:3}
\end{figure}

\section{Results}
Throughout the NEAIMD-EPO simulation, we can obtain the spatial distribution and the dynamical evolution of the electrostatic potential $U(R)$, where $R$ represents the ion position\citep{Yue}. Fig.~\ref{fig:1}(c) presents the statistical results of the effective amplitudes of the EPO in NEAIMD for $\epsilon-$iron lattice in space. Fig.~\ref{fig:2} exhibits the evolution of system pressure and temperature with respect to simulation time for the entire NEAIMD run as presented in Fig.~\ref{fig:1}. To intuitively demonstrate the dynamic evolution of spatial EPO, we plot the spatial EPO with simulation time in Fig.~\ref{fig:3}. We can observe the local electric-field variation between neighboring atom layers, and the local-field directions continually changing with time. The local electric-field variation will drive the collective vibration of whole free electrons in it, and then in momentum space the free electrons near the Fermi surface could be thermally excited. The local electric-filed oscillation will provides additional kinetic energy to the TEEs through the collective oscillating process of the free electrons in it\citep{Yue}. Fig.~\ref{fig:1}(c) shows the distribution of the effective amplitude of EPO ($\overline{U}_{EPO}(l)$) per atoms-layer along the $\nabla \vec{T}$, where $l$ is the index for atom layers. Moreover, the amplitude distribution of EPO indicates that how the thermal kinetic energy of TEEs is divided in space. More information of the average effective amplitude of EPO of $\epsilon-$iron at different Earth's core conditions can be found in Sec.~2.3 of SI. Here, we calculate $\overline{U}_{EPO}(l)$ using the root mean square (RMS) method\citep{Yue}:
\begin{eqnarray}
\overline{U}_{EPO}(l)=\frac{1}{N_{al}}\sum_{j=1}^{N_{al}}\sqrt{\frac{1}{n_{steps}}\sum_{t_{i}}^{n_{steps}}(U_{j}(t_{i})-\overline{U}_{j})^2},
\end{eqnarray} 
where $n_{steps}$ is the simulation steps, $U_{j}(t_{i})$ is the value of the electrostatic potential $U$ for atom $j$ in a specific layer at the specific time step $t_{i}$, and $\overline{U}_{j}$ is the average value of $U_{j}(t_{i})$. Then, the electronic heat flux $\vec{J}_{el}$ according to the kinetic energy of TEEs between two adjacent atom layers can be defined as\citep{Yue}: 
\begin{eqnarray}
\vec{J}_{el}&=&-\frac{n(e)\cdot e\cdot n_{steps}}{S\cdot t}\frac{\partial\overline{U}_{EPO}(l)}{\partial N_{l}},
\end{eqnarray}
where $S$ represents the cross-sectional area of the simulation model, $t$ is simulation time, $n(e)$ is the valance electrons number per atoms layer, and $\frac{\partial\overline{U}_{EPO}(l)}{\partial N_{l}}$ is the gradient in space of the effective amplitude value of EPO via the linear fitting for $\overline{U}_{EPO}(l)$ with the atom layer index $N_{l}$ in Fig.~\ref{fig:1}(c) (excluding heat source and heat sink layers). See the theoretical details in Sec.~2.1 of SI. Finally, based on the Fourier\rq s law we can calculate $\kappa_{el}$ of $\epsilon-$iron as\citep{Yue}:
\begin{eqnarray}
\kappa_{el}=\frac{n(e)\cdot e \cdot n_{steps}}{\nabla T \cdot S\cdot t}\frac{\partial\overline{U}_{EPO}(l)}{\partial N_{l}},
\end{eqnarray}
where $\nabla T$ can be obtained via the linear fitting for temperature profile. The data of a representative case are shown in Fig.~\ref{fig:1}(b). 

\begin{table}
\caption{
The electronic ($\kappa_{el}$) and phononic ($\kappa_{ph}$) part of thermal conductivity of $\epsilon-$iron at Earth's core conditions obtained from our NEAIMD-EPO method. Some previous theoretical (based on Kubo-Greenwood method) and experimental results (based on the Wiedemann-Franz law and direct measurement) are also given for comparison. The unit of thermal conductivities is $\rm W/(mK)$.
}
\label{tb-1}
\begin{tabular}{*{5}{|c}|}
\hline
Condition & P~(GPa) & ~~T~(K)~~ & $\kappa_{el}$ & $\kappa_{ph}$  \\ 
\hline
Solid (CMB) & 136 & 3741 & 198 [this work] & 11 \\
Solid (CMB) & 137 & 3916 & 184 [this work] & 13 \\
\hline
Solid (ICB) & 316 & 5289 & 199 [this work] & 13 \\
Solid (ICB) & 324 & 5905 & 167 [this work] & 11 \\
Solid (ICB) & 327 & 5544 & 182 [this work] & 11 \\
Solid (ICB) & 327 & 5835 & 174 [this work] & 12 \\
Solid (ICB) & 331 & 6378 & 179 [this work] & 12 \\
Solid (ICB) & 361 & 5435 & 198 [this work] & 15 \\
Solid (ICB) & 362 & 5662 & 201 [this work] & 13 \\
Solid (ICB) & 365 & 6239 & 189 [this work] & 13 \\
\hline
\hline
Solid (ICB) & 329 & 5500 & 286 \citep{Pozzo-EPSL} & \\
Solid (ICB) & 364 & 5500 & 297 \citep{Pozzo-EPSL} & \\
Solid (ICB) & 329 & 5700 & 294 \citep{Pozzo-EPSL} & \\
Solid (ICB) & 364 & 5700 & 307 \citep{Pozzo-EPSL} & \\
\hline
Solid (ICB) & $\rm \sim330$ & 6000 & $\rm \sim190$ \citep{Pourovskii} & \\
\hline
Liquid (CMB) & 134 & 4700 & 160 \citep{Pozzo-nature} & \\
Solid (ICB) & 320 & 6250 & 243 \citep{Pozzo-nature} & \\
Solid (ICB) & 329 & 6350 & 247 \citep{Pozzo-nature} &\\
Solid (ICB) & 339 & 6435 & 248.5 \citep{Pozzo-nature} & \\
\hline
Solid (ICB) & 360 & 8000 & 270 \citep{Koker-pnas} &\\
\hline
Solid (ICB) & 342 & 6000 & 210 \citep{Koker-pnas} &\\
\hline
Liquid (CMB) & 140 & 3750 & 226 measured by \citep{Ohta} (WFL) &\\
\hline
Liquid (CMB) & 136 & 3750 & 67-145 measured by \citep{Seagle} (WFL) & \\
\hline
Liquid (CMB) & 136 & 3800 & 33 directly measured by \citep{Konopkova} & \\
Solid (ICB) & 330 & 5600 & 46 directly measured by \citep{Konopkova} & \\
\hline 
\end{tabular}
\end{table}

\begin{figure}[th]
\includegraphics[width=1.0\linewidth,clip]{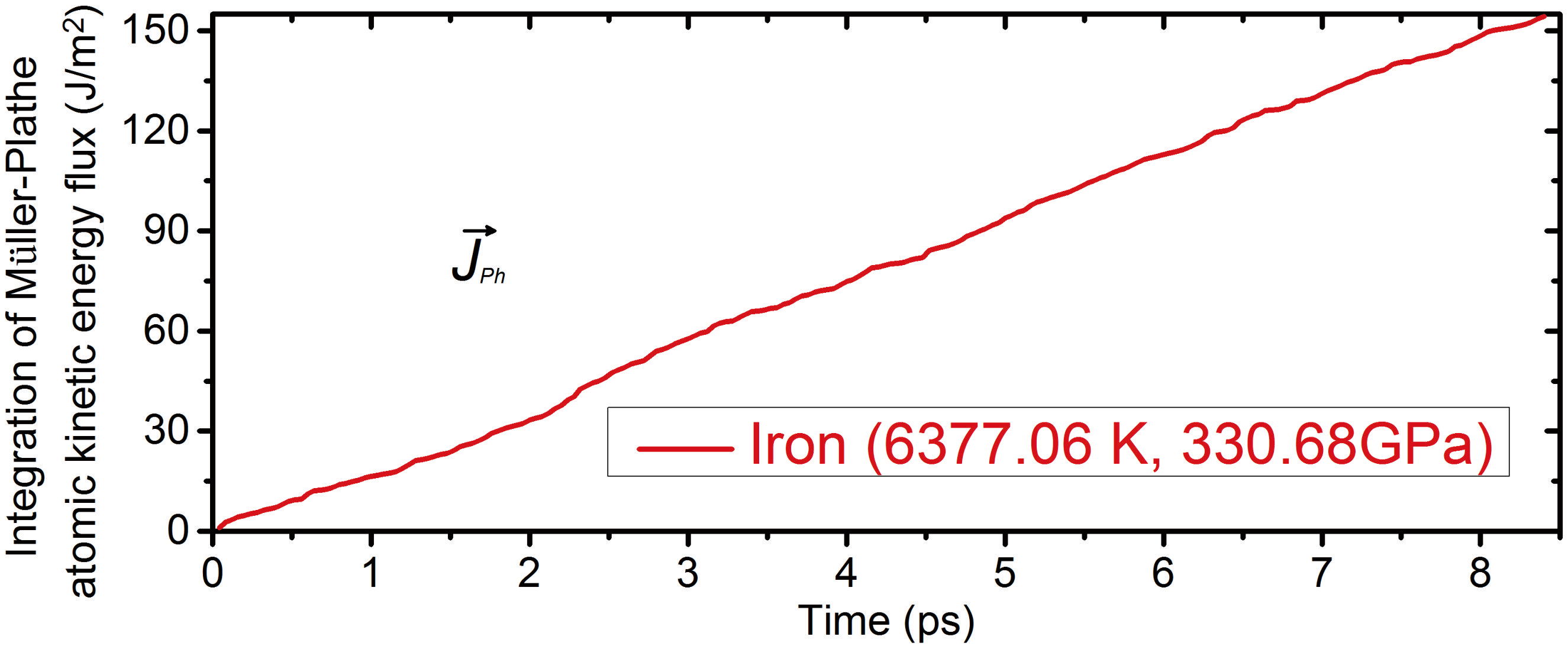}
\caption{The corresponding integration of the heat flux from the atomic kinetic energy part with time for the case shown in Figs.~\ref{fig:1}. Here, the slope defines the phonon heat flux which can be used to calculate phonon thermal conductivity.}
\label{fig:4}
\end{figure}

Additionally, via integrating the atomic heat flux in the M\"uller-Plathe algorithm, as shown in Fig.~\ref{fig:4}, we can estimate the $\kappa_{ph}$ of $\epsilon-$iron at the Earth's core conditions simultaneously. Here, we adopt an approximation usually used in statistical physics: with sufficient simulation time, the time-average values of $\kappa_{el}$ and $\kappa_{ph}$ should be equal to the ensemble-average values. We have performed the test for the convergence of simulation time and system size. Because of the size-effect limitation, the NEAIMD method will underestimate the $\kappa_{ph}$. We also implemented the phonon Boltzmann transport equation (BTE) method\citep{shengbte01,shengbte02,shengbte03} to calculate the $\kappa_{ph}$ of $\epsilon-$iron for comparison. However, because that the fourth- and higher order anharmonicity of phonons will become very important at high temperature\citep{Yue-RRA}, the phonon BTE method would overestimate the $\kappa_{ph}$. The details of discussion for $\kappa_{ph}$ can be found in Sec.~2.8 of SI.

\begin{figure*}
\includegraphics[width=0.9\linewidth,clip]{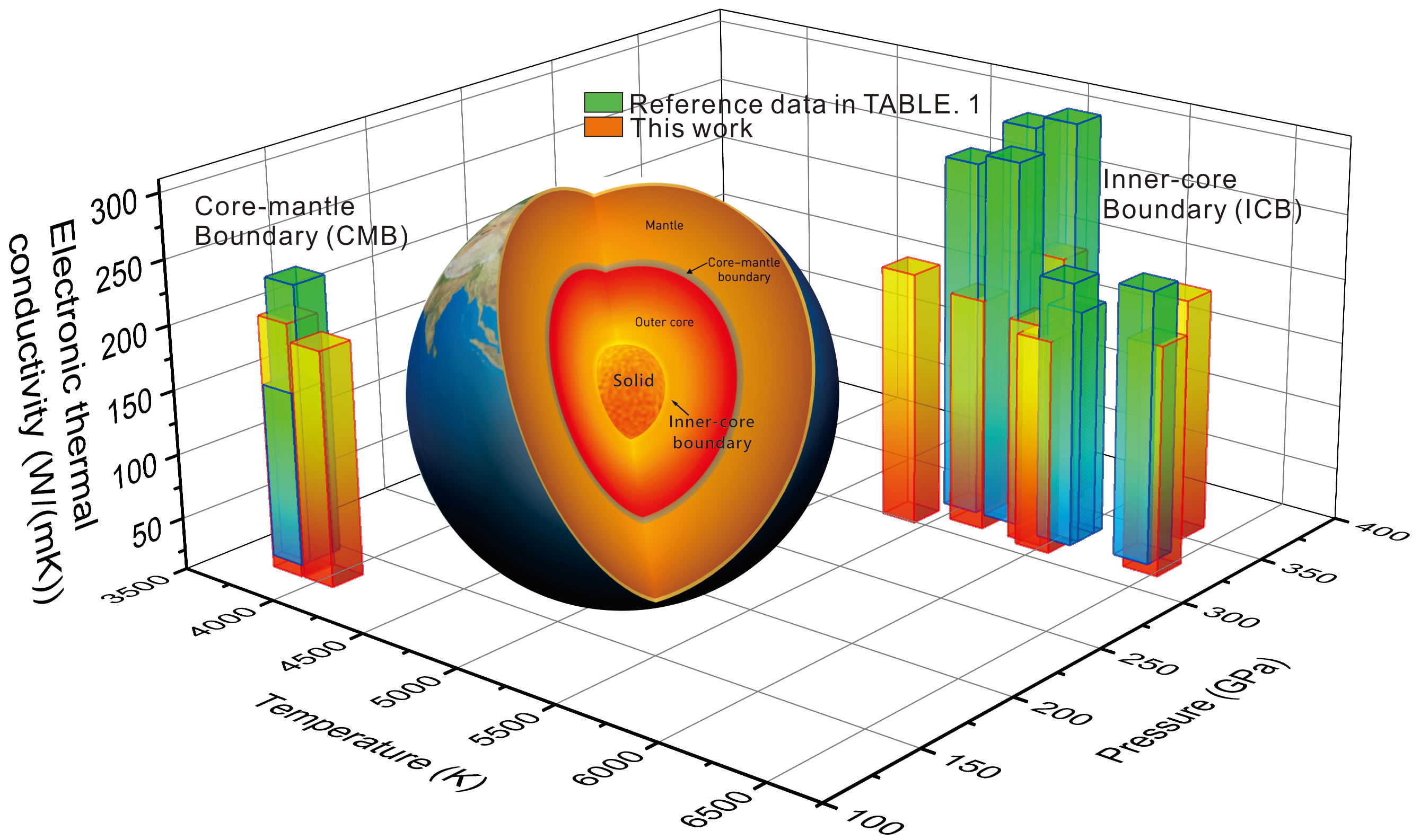}
\caption{The electronic thermal conductivities of $\epsilon-$iron under the Earth's core conditions (CMB and ICB). The red and green bars represent the results from NEAIMD-EPO method and previous studies, respectively.}
\label{fig:5}
\end{figure*}

The main simulation results are presented in TABLE.~\ref{tb-1} and Fig.~\ref{fig:5}. We also calculated the errors of $\kappa_{el}$ in TABLE.~\ref{tb-1} via the $\kappa_{el}$ expression and error propagation theory\citep{Ku}, which mainly stems from the gradient of $\overline{U}_{EPO}(l)$ and $\nabla T$\citep{Yue}. The errors of $\kappa_{el}$ and  more details are provided in Sec.~2.9 of SI. The non-linear effect analysis from temperature and simulation size for $\kappa_{el}$ of $\epsilon-$iron are shown in Sec.~2.4 of SI. The errors of $\kappa_{ph}$ estimated in TABLE.~\ref{tb-1} mainly originate from the the nonlinearity of the temperature gradient $\nabla T$. 
To intuitively see the $\kappa_{el}$ results from NEAIMD-EPO method, in Fig.~\ref{fig:5} we also plot the $\kappa_{el}$ of $\epsilon-$iron from this work and previous theoretical and experimental data in Fig.~\ref{fig:5} with different color bars in 3D style. 

Our results at ICB conditions are comparable to the previous theoretical works from Kubo-Greenwood (KG) formula\citep{Pozzo-EPSL, Koker-pnas}.
Especially, they are well consistent with the recent theoretical research by Pourvoskii, et al.\citep{Pourovskii}. In the research of Pozzo et al.\citep{Pozzo-EPSL}, they reported that the $\kappa_{el}$ of pure solid $\epsilon$-iron at $\rm 5500~K$ and $\rm 329~GPa$ is $\rm 286~W/(mK)$, which is larger than the value $\kappa_{el}$ at $\rm 6000~K$ and $\rm 342~GPa$ reported by Koker et al\citep{Koker-pnas}. In research by Pourvoskii et al.\citep{Pourovskii}, they demonstrate that the electron-electron scattering effect can be comparable to the electron-phonon scattering. They treat the $\epsilon-$iron as strongly correlated electrons system ``Fermi-liquids" and consider both the electron-phonon and electron-electron scattering. The Fermi-liquid model seems to be more accurate to describe the thermal transport of $\epsilon-$iron under ICB conditions. However, as the many-body problem is too complicated, there could still lack of some complicated interactions and scattering mechanism of electrons compared with the real situation. Then, they modify the WFL and give a prediction of $\kappa_{el}$ of perfect $\epsilon-$iron crystal phase at Earth's core conditions as $\rm \sim 190~W/(mK)$ at $\rm \sim 330~GPa$ and $\rm \sim 6000~K$. This value is smaller than previous theoretical predictions by Pozzo et al.\citep{Pozzo-EPSL} and Koker et al.\citep{Koker-pnas}. In addition, our $\kappa_{el}$ results of $\epsilon-$iron at Earth's core conditions are also compatible with some recent experimental measurements\citep{Ohta,Seagle}.


In comparison, our NEAIMD-EPO method\citep{Yue} can directly simulate the thermal transport process in pure metals. In principle, our method inherently includes all electron-electron and electron-phonon scattering process. Thus, the NEAIMD-EPO method can theoretically reveal the real physical picture of the thermal transport in $\epsilon-$iron at the Earth's core conditions. Our results in TABLE.\ref{tb-1} ($\rm 179~W/(mK)$ at $\rm 331~GPa$, $\rm 6378~K$) are very close to the value of $\rm \sim 190~W/(mK)$ by Pourvoskii, et al.\citep{Pourovskii} under the similar temperature and pressure conditions. The slight difference can be attributed to two facts: (1) the temperature effect; (2) since our method includes all the scatterings of electrons theoretically, our $\kappa_{el}$ result ($\rm 179~W/(mK)$) is slightly lower than the $\kappa_{el}$ ($\rm \sim 190~W/(mK)$) from Pourovskii et al.\citep{Pourovskii}. Our results of $\kappa_{el}$ of $\epsilon-$iron at ICB and CMB conditions are also consistent with one of experimental works by Ohta, et al.\citep{Ohta}, which confirm the high thermal conductivity of the Earth's core. In addition, another advantage of our NEAIMD-EPO method is that, under the NEAIMD-EPO framework we do not need to figure out the electron collision lifetime $\tau_{el}$, which is usually obtained by the Matthiessen's rule and mainly originates from the electron-phonon interaction (EPI) and is generally very difficult and complicated to accurately evaluate.

\section{Conclusion}
To conclude, based on first-principles calculations, we use our newly developed methodology of direct non-equilibrium $ab~initio$ molecular dynamics simulation coupled with the electrostatic potential oscillation (named as ``NEAIMD-EPO method") to simulate the real thermal transport process of $\epsilon-$iron under the Earth's core conditions. The NEAIMD-EPO method can describe the thermal transport in the Earth's core conditions in a new physical picture. The method intrinsically includes all electron-electron scatterings and electron-phonon scatterings, and can predict the electronic and phononic thermal conductivity simultaneously. The NEAIMD-EPO method is parameter free and shows robustness in the simulations of the thermal transport process of $\epsilon-$iron at the Earth's core conditions. Our simulation results are comparable to and are consistent with some recent theoretical and experimental works and confirm that the thermal conductivity of $\epsilon-$iron in the Earth's core is high. The NEAIMD-EPO method provides a new approach to study the thermal transport properties of metals in extreme conditions. We also expect that our work could be helpful for clarifying the physical nature of the thermodynamic evolution in the Earth's core.

\subparagraph{Data availability}
Authors can confirm that all relevant data are included in the paper and the supplementary information files.


%

\acknowledgments
All simulation details can be found in the Supplement Information. The authors thank Dr. Stephen Stackhouse (University of Leeds) for fruitful and insightful discussions. The authors also gratefully acknowledge the computing time granted by the John von Neumann Institute for Computing (NIC) and provided on the supercomputer JURECA at J\"ulich Supercomputing Centre (JSC) (Project No. JHPC25 and JHPC46). M.H. acknowledges the start-up fund from the University of South Carolina.

\section{Author contributions}
S.-Y.Y. performed the simulations and analysis. M.H. supervised the project. S.-Y.Y. and M.H. write the paper.

\section{Additional information}

\subparagraph{Competing interests}
The authors declare no competing financial interests.

\bibliography{main}

\end{document}